\documentstyle[aps,twocolumn,epsfig]{revtex}

\title{Macroscopic quantum state reduction: uniting Bose-Einstin condensates
by interference measurements.}

\author{Klaus M\o lmer \\
{\small
  Institute of Physics and Astronomy, University of Aarhus}\\
{\small DK-8000 \AA rhus C, Denmark}}
\date{\today}

\begin{document}
\draft

\maketitle

\begin{abstract}
Two independently prepared condensates can be combined into a single
larger condensate by detection of  their relative phase in an intereference
measurement. 
\end{abstract}


It is a highly relevant and desirable experimental task to
merge condensed atoms  to form large Bose-Einstein condensates.
This, in principle will allow
us to load a condensate with atoms which are trapped and 
cooled  in a spatial
region which does not interfere with the condensate dynamics,
and if this is possible,
we can produce much larger condensates than the ones studied today,
and we may operate atom lasers in a continuous manner. 
Transport of cold atoms  has been demonstrated by different means,
involving current carrying wires \cite{denschlag}, 
far-off resonant dipole potentials \cite{renn},
and chains of magnets \cite{greiner}, but
it is not evident how one would merge the macroscopic
condensate wave functions to form a single larger condensate.

Let us assume that two orthogonal single particle wave functions 
$\phi_a$ and $\phi_b$
are macroscopically populated with $n_a$ and $n_b$ atoms. The state
vector of the entire system is thus the tensor product state
$\phi_a^{n_a}\otimes\phi_b^{n_b}$. To unite the atoms into a single
condensate means to drive this state into a state of the form $\phi^n$,
where $n=n_a+n_b$, and $\phi$ is a single particle wave function.
Both the initial and the final tensor product states belong to the Hilbert 
space of the full many body problem, and there  exists
a unitary evolution that unites the condensates. 
For example, one may split a condensate into two
components so slowly that the state vector follows the
minimum energy state adiabatically, and if the atoms repel each other,
the ground state of the split condensate is the product of two
number states \cite{menotti}. 
Running this adiabatic process backwards will therefore combine 
two condensates into a single one.
If, however, the atoms do not interact with each other external
manipulation of the trapping potential will only cause
a unitary evolution of the two single particle wave function 
$\phi_a$ and $\phi_b$; they will remain orthogonal, and it will 
not be possible to unite the two components into a single condensate. 
A recent idea to break unitarity 
couples the system to an ancillary dissipative
degree of freedom in the form of a lossy optical cavity field.
In analogy with proposals for
cavity assisted laser cooling of atoms, Jaksch et al, show that the
unitary single atom dynamics can be avoided, and the condensates are
"cooled" together \cite{jaksch}.

We suggest that the easiest and most natural way to unite condensates
is by quantum measurements. 
The wave function is the mathematical representation of our knowledge
about a physical system. When we acquire more information by means of 
measurements, the wave function changes, and we can use this to 
prepare states which would be difficult to prepare by other methods.
In few-body systems, detection of particles belonging to different
Einstein-Podolsky-Rosen photon pairs have enabled experimentalists to
entangle particles which have never met \cite{zeilinger}.
Recently, spin squeezed states of atoms have been prepared
\cite{kuzmich}, and  methods
to entangle separate ensembles of atoms have been proposed \cite{polzik}
making use of state reduction.  In order to unite two condensates we
only have to measure the,
{\it a priori} undefined, relative phase between the two condensates.

Few years ago, theorists and experimentalists raised the issue of 
phase of the macroscopic wave function of a dilute atomic gas.
Theoretically, it takes a breaking of the $U(1)$ symmetry associated
with particle number conservation to make such a phase appear naturally,
and hence the experimental observations of interference
between two condensate components \cite{davis} was met with large interest.
Similar interest was received by theoretical works, pioneered by
Javanainen et al \cite{javanainen,javanainen2}, showing that, indeed, the
macroscopic interference observed in experiments with merging atomic samples,
can be understood fully in terms of quantum mechanical state
reduction occuring during the first random detection events.
Two independent condensates acquire a random relative phase while the
interference signal is measured, as illustrated in 
\cite{javanainen} by  a numerical simulation of two overlapping 
number state  condensates arriving at a position sensitive detector. 
The effects of interactions were further analyzed in \cite{naraschewski} 
to model more precisely the experimental results. 

%

When the relative phase
between two condensates is established, the two condensates effectively
join and become one single condensate. This is in fact also the idea in
Ref.\cite{jaksch}, where dissipation ``cools" the relative phase, but
so far it was never an issue in the discussion of interference
measurements.

In a simple model analyzed by Castin and Dalibard
\cite{castin} atoms leave their respective 
separate condensates to be detected in one of two superposition
states, unable to 
distinguish from which condensate component the atom originated.
We write $|n_a,n_b\rangle$ for  the initial separable product state 
of the atoms.
The detection and annihilation
of a single atom in the state $\frac{1}{\sqrt{2}}(\phi_a+\phi_b)$ leaves
the remaining atoms in an entangled  state 
$(\sqrt{n_a}|n_a-1,n_b\rangle +\sqrt{n_b}
|n_a,n_b-1\rangle)/\sqrt{n_a+n_b}$. 
If we asume that $n_a \simeq n_b$ and both are large numbers, we 
observe that 
there is now a 75 \% probability that the 
subsequent atom is detected  in the same state and
only 25 \% probability that it is detected in the state
$\frac{1}{\sqrt{2}}(\phi_a-\phi_b)$ \cite{castin}.
A 75 \% detection probability means that the ``detector mode"
$\frac{1}{\sqrt{2}}(\phi_a+\phi_b)$ 
is populated by 75\% of the atoms. I.e., the detection of
only one atom out of, possibly, a macroscopic number raises the
''condensate fraction" from 50 \% to 75 \%, where by condensate
fraction we refer to the population of the mostly occupied singe 
particle wave function.  Castin and Dalibard investigated the
effects of further detection events, and they found that after $k$
detection events the absolute value of
the relative phase $\theta$ between the two condensates is determined 
up to a width of order $1/\sqrt{k}$ \cite{castin}.

The detection scheme of Castin and Dalibard is 
able to identify only the absolute value of the phase, 
but as shown in a similar analysis for light by the present author 
\cite{molmer}, a difference in frequency, i.e.,
energy per particle between the two components, causes a phase rotation
which breaks the symmetry of phase values $\pm \theta$, and the 
relative phase of the components, including the value
of the sign, is well determined.
Precise simulations were carried out for photons in \cite{molmer},
but we note that the
conclusions obtained for optical fields can be applied with no changes 
to non-interating atomic systems.
Letting $a^{\dagger}(a)$ and
$b^{\dagger}(b)$ denote creation (annihilation) operators for particles in
the two separate modes $\phi_a$ and $\phi_b$, the branching ratio 
for detection in the two superposition states
$\frac{1}{\sqrt{2}}(\phi_a\pm \phi_b)$
is controlled by the expectation values
$\langle a^{\dagger} a + b^{\dagger}b 
\pm (a^{\dagger}b +b^{\dagger}a)\rangle$, 
which at each detection even must be 
evaluated in the state of the
system.  Initially $\langle a^{\dagger} b\rangle=0$, but
as shown in Fig. 3. of Ref. \cite{molmer},
after very few detection events $\langle a^{\dagger} b\rangle$ 
becomes a harmonically varying
function with absolute value equal to the number of atoms in each field
mode. 

There is of course a direct formal relationship between the
interference signal and the formation of a single condensate.
The condensate fraction is determined by the largest eigenvalue of
the one-body density matrix for the atoms, which in
the present case  is a 2 by 2 matrix,
corresponding to the two mode functions $\phi_a$ and $\phi_b$,
\begin{eqnarray}
\rho_{(1)} = \left( \begin{array}{cc} 
\langle a^{\dagger}a \rangle & \langle a^{\dagger}b \rangle \\ 
\langle b^{\dagger}a \rangle & \langle b^{\dagger}b \rangle \\ 
\end{array} \right).
\label{rho1}
\end{eqnarray}
For $\langle a^{\dagger}a \rangle = \langle b^{\dagger}b
\rangle$, the largest eigenvalue of $\rho_{(1)}$ equals
$\langle a^{\dagger}a \rangle + |\langle a^{\dagger}b \rangle |$
approaching the total number of atoms in the limit of unit visibility 
$2|\langle a^{\dagger}b\rangle|/\langle a^{\dagger}a+b^{\dagger}b\rangle
= 1$. If the two components have unequal populations there is a
well-known reduction of the interference signal, but the maximum
attainable interference still implies a unit condensate fraction
(in a state with different amplitudes on $\phi_a$ and $\phi_b$).

The theoretical analyses \cite{javanainen,naraschewski,castin,molmer}
all apply a scheme, where the particles (atoms or photons) are absorbed  
during detection.
It is clear from the simulations and the analytical estimates, that 
only relatively few particles  
have to be sacrificed to establish the relative phase
and the single mode character of the united condensate. Particle
detection thus offers a realistic practical proposal to unite 
condensates. It is also
possible to detect the interference 
by optical imaging, revealing in a non-destructive manner the
interference between 
the overlapping condensates. After transmission
of a suficiently large number of photons to produce a full contrast
optical interference pattern, the atoms are known to populate the
spatial wave function measured. 

The macroscopic wave function produced is of the form
$\phi^n$, where $\phi$ is a superposition of $\phi_a$ and $\phi_b$
with a phase difference which 
is revealed by the detection process.
It is now a matter of wave function
engineering to transform this wave function into the one appropriate
for further experiments, e.g, one fitting 
a trap potential. If, for example, $\phi_a$ and $\phi_b$ are
extended state with well defined momenta $0$ and $2\hbar k$, after 
establishing the relative phase, a controlled Bragg pulse 
induced by a moving standing optical wave
\cite{denschlag2} can drive
the entire condensate into just one of the two momentum states.
Also, atoms with two different internal states can be spatially
overlapped, and the detection may be used to produce a
macroscopically populated superposition state, which may subsequently
be rotated into any desired internal state.
In situations where one wants to load a larger trapped condensate,
we assume that the experimentalist masters the coupling of the
incident transport mode to the trapped condensate mode, and future
loading makes use of precisely the same operation, with a 
phase parameter to be adjusted according to the outcome of 
interference measurements.

In conclusion, we have pointed out, that by detecting the interference
between two condensates, one is not only setting up the relative phase,
responsible for the interference, one is actually 
producing a state of the system with macroscopic population of 
one single-particle wave function. Other methods do exist
for the same task, but doing it by interference measurements
seems experimentally feasible, in fact to have already been done in
experiments. 
We intentionally disregarded the role of interactions. 
Our scheme works explicitly in the absense of 
collisional interactions in the condensate, and we suggest to operate
the process in that regime.  Interactions
cause dephasing of the relative phase between condensates, and 
in cases where they cannot be controllably removed, they may still
be disregarded if interference  measurements occur on a faster time 
scale as analyzed in \cite{castin}.

\end{document}